# HANDOVER ADAPTATION FOR DYNAMIC LOAD BALANCING IN 3GPP LONG TERM EVOLUTION SYSTEMS

Ridha Nasri[*], Zwi Altman

**Abstract**
*The long-Term Evolution (LTE) of the 3GPP (3$^{rd}$ Generation Partnership Project) radio access network is in early stage of specification. Self-tuning and self-optimisation algorithms are currently studied with the aim of enriching the LTE standard. This paper investigates auto-tuning of LTE mobility algorithm. The auto-tuning is carried out by adapting handover parameters of each base station according to its radio load and the load of its adjacent cells. The auto-tuning alleviates cell congestion and balances the traffic and the load between cells by handing off mobiles close to the cell border from the congested cell to its neighbouring cells. Simulation results show that the auto-tuning process brings an important gain in both call admission rate and user throughput.*

## 1. Introduction

3GPP organization is defining the requirements for an evolved UTRAN (e-UTRAN: UMTS Terrestrial Radio Access Network) and is in the beginning of the specification stage. The evolution of 3G systems is referred to the 3GPP Long Term Evolution (LTE) systems. LTE system, called sometimes super 3G, is expected to offer a spectral efficiency between 2 to 3 times bigger than 3GPP release 6. It will provide up to 100Mbit/s for 20 MHz of spectral bandwidth. Both the radio and the core network parts of the LTE technology are impacted: The system architecture is more decentralized; The RNC (radio network controller) present in the 3G systems is removed; and the Radio Resource Management (RRM) functionalities are moved to an "upgraded" base station called Evolved Node B (eNB) [1, 4].

To improve spectral efficiency, a new radio access technology is introduced based on the OFDMA (Orthogonal Frequency Division Multiple Access), and will be used in the downlink instead of the current CDMA (Code Division Multiple Access) technology. In OFDMA systems, the total bandwidth is split into many sub-carriers [2]. Each sub-carrier is exclusively assigned to only one user, eliminating the intra-cell interference, the most limiting factor in CDMA systems. Furthermore, OFDMA allows a high flexibility in resource allocation and scalability in terms of utilized bandwidths. The LTE system will support different bandwidth allocations going from 1.25 MHz and up to 20 MHz [4].

The LTE working group 3GPP TSG-RAN WG3 specifically states that the work on radio interface protocol architectures should include mobility solutions, self-configuration and self-optimisation of the e-UTRAN nodes. In the first phase of Optimisation/Adaptation, the neighbour list optimisation, coverage and capacity control have been proposed [5]. The study of mobility parameter auto-tuning has been identified as a relevant case study of self-configuration. The aim of the case study is to evaluate the contribution of this functionality to the LTE performance.

---

[*] R. Nasri and Z. Altman are with Orange Labs, 38-40 rue du Général Leclerc, 92794 Issy les Moulineaux, France. ridha.nasri@orange.com, zwi.altman@orange.com

This paper investigates the performance of the auto-tuning process of LTE hard handover (HO) parameters. Based on the loads of eNBs, the auto-tuning is performed by dynamically adapting the HO margins. The HO margin is the main parameter that governs the HO algorithm between two eNBs. The auto-tuning of soft handover parameters in a WCDMA (Wideband CDMA) network has been shown particularly effective for traffic balancing and for increasing network capacity [6, 7].

The rest of this paper is organized as follows: In section II, the system model including inter-cell interference is described. Section III deals with the LTE hard handover algorithm and its auto-tuning. In section IV, simulation results are presented for the auto-tuning case and are compared with the fixed handover margin case. Section V concludes the paper.

## 2. Model and definition

### 2.1. System Model

Consider an e-UTRAN radio access network composed of a set of eNBs (see Figure 1). Each eNB covers a geographical area (called cell) and serves a number of users in each service class with a target quality of service (e.g. blocking and dropping rates, minimum throughput etc.). Continuous coverage over more than two eNBs is achieved by hard handover mechanism (soft handover was abandoned by 3GPP for the LTE standard).
Each user uses a given number of sub-carriers, and unlike the 3G-CDMA systems, there is no intra-cellular interference. The maximum resource in an eNB is characterized by the total number of physical resource blocks (PRB) available in the cell at each time slot, named also TTI (Time Transmit Interval). Each PRB comprises 12 sub-carriers of 15 KHz each one. We define the load of an eNB $\chi$ as the ratio between the occupied PRB and the total amount of PRB.

An *Ukumara-Hata* propagation model is used in the 2 GHz band. The attenuation $L$ due to the propagation is given by $L = l_o d^\gamma \zeta$, where $l_o$ is a constant depending on the used frequency band, d is the distance between the eNB and the mobile, $\gamma$ is the path loss exponent and $\zeta$ is a log-normal random variable with zero mean and standard deviation $\sigma$ representing shadowing losses.

### 2.2. Interference model

In LTE systems, the same frequency band can be used by all eNBs. This generates inter-cell interference which limits the performance of the LTE system. In the downlink for instance, inter-cell interference occurs at a mobile station when a nearby eNB transmits data over a sub-carrier used by its serving eNB. The interference intensity depends on user locations, frequency reuse factor and loads of interfering cells. For instance, with a reuse factor equals to 1, very low cell-edge performances are achieved whereas for reuse factor higher than 1, the cell-edge problem is resolved but resource limitation occurs.
The interference model used in the present study is the following:

Let Λ denote the interference matrix of LTE system, where the coefficient Λ(*i,j*) equals to 1 if cells *i* and *j* use the same band and zero otherwise.

The average downlink interference per sub-carrier $I_{e,m}$ perceived by a mobile station *m* connected to an eNB *e* is computed by:

$$I_{e,m} = \sum_{k \neq e} \Lambda(e,k) \chi_k \frac{P_k G_{k,m}}{L_{k,m}} \quad (1)$$

where $P_k$ is the downlink transmit power per sub-carrier of the eNB *k*. $G_{k,m}$ and $L_{k,m}$ are respectively the antenna gain and the path loss between eNB *k* and the mobile station *m*. The factor $\chi_k$ is the probability that the same sub-carrier used by the mobile *m* is used in the same time by another mobile connected to the eNB *k*. The factor $\chi_k$ is exactly the load of the eNB *k* when system performances are taken over a more or less long period (~15mn). The sum in equation (1) is over all eNBs in the network except eNB *e*. The downlink quality can be measured by the Signal to Interference plus Noise Ratio (SINR)

$$SINR_{e,m} = \frac{P_e G_{e,m}}{L_{e,m}(I_{e,m} + N_{th})} \quad (2)$$

where $N_{th}$ is the thermal noise per sub-carrier.

In LTE system, an adaptive modulation and coding scheme will be used [2, 3]. So, the choice of the modulation depends on the value of the SINR through the perceived Bloc Error Rate (*BLER*). The decrease of the *SINR* will increase the *BLER*, forcing the eNB to use a more robust (less frequency efficient) modulation. This may have negative impacts on the communication quality. For instance, a lower efficiency modulation results in a lower throughput and a larger transfer time for elastic data connections. In the present paper, the throughput per PRB for each user is determined by a link level curve that maps the SINR to the throughput. The used link curve is obtained from a link level simulator tool developed internally in Orange Labs. The user physical throughput is $N_m$ times the throughput per PRP, where $N_m$ represents the number of PRP allocated to the user *m*.

## 3. LTE Handover algorithm and its auto-tuning

Unlike 3G-CDMA systems, mobility in LTE will be based on hard handover rather than on soft handover [7]. The mechanism of hard handover has been used in 2[nd] generation GSM networks and the basic concept will be likely implemented in LTE systems. In this paper we consider power budget based handover. This handover is based on the comparison of the received signal strength from the serving cell and from the neighbouring cells.

### 3.1. Call admission control and handover algorithm

In order to study the performance of the auto-tuning of LTE hard handover, some assumptions for the call admission control (CAC) and resource allocation are made.
In the CAC algorithm, a user can be admitted to the network only when the following conditions are fulfilled:

- Good signal strength: The mobile selects the eNB that offers the maximum signal. If this signal is lower than a specified threshold then the mobile is blocked because of coverage shortage. This condition is similar to a selection/reselection algorithm when mobile is in idle mode.
- Resource availability in the selected eNB: The mobile can be granted physical resources in terms of PRP between a minimum and maximum threshold. When the first condition is satisfied, the eNB checks the available resource. If the latter is lower than the minimum resource threshold, the call is blocked.

Hard handover is performed in this study using a similar algorithm to the one used in GSM systems. While in communication, the mobile periodically measures the received power from its serving eNB and from the neighbouring eNBs. The mobile, initially connected to a cell *e*, triggers a handover to a new cell *k* if the following conditions are satisfied:

- The Power Budget Quantity (PBQ) is higher than the handover margin: a mobile *m* connected to a eNB *e* triggers a handover to an adjacent eNB *k* if:

$$PBQ = P_k^* - P_e^* \geq HM(e,k) + Hysteresis \qquad (3)$$

where $P_k^*$ is the received power from the eNB *k* expressed in dBm, $HM(e,k)$ is the handover margin between eNB *e* and *k*, the *Hysteresis* is a constant independent of the eNBs and mobile stations and is fixed in this study to 0.
- The received power from the target eNB must be higher than a threshold. This is the same condition as in the CAC process but here the threshold is not necessary the same as for admission control.
- Enough resources are available in the target eNB.

According to (3), adapting the handover margin can delay or advance the establishment of new radio connections with the target cell, according to the load of both old and target cells.

### 3.2. Auto-tuning of handover margin

The auto-tuning aims at dynamically adapting handover margins between cells as a function of their loads, to optimize network performances. Each coefficient of the matrix *HM* governs the traffic flows between two cells. The coefficient $HM(e,k)$ only depends on the difference of the load of cell *e* and *k*. Define the handover margin matrix *HM* as

$$HM(e,k) = f(\chi_e - \chi_k) \qquad (4)$$

*Theorem*
For the dynamic load balancing, the function *f* must satisfy:
1) *f* is a decreasing function from the compact interval [-1,1] to [$HM_{min}$, $HM_{max}$],
2) $f(x) + f(-x) = 2f(0), \forall x \in [-1,1]$

where $HM_{min}$ and $HM_{max}$ are respectively the minimum and the maximum values of the handover margin. $f(0)$ is the value of the planned handover margin since the planning process assumes the uniformity of the cell loads.

*Proof*
The first condition is evident. For the second condition, let $\chi_e$ be the load of cell *e*, $\chi_k$ be the load of its neighbour cell *k* and $x = \chi_e - \chi_k$, the auto-tuning algorithm does not perform well if cell *e* and *k* increase or decrease simultaneously their handover margin:

i) The auto-tuning system crashes if $HM(e,k) = f(x)$ and $HM(k,e) = f(-x)$ exceed $f(0)$; thus $f(x) + f(-x) > 2f(0)$

ii) The second crash is when $HM(e,k)$ and $HM(k,e)$ go at the same time under $f(0)$; thus $f(x) + f(-x) < 2f(0)$

By combining (*i*) and (*ii*), the auto-tuning crashes when $2f(0) < f(x) + f(-x) < 2f(0)$. So, the auto-tuning algorithm performs well only with the condition $f(x) + f(-x) = 2f(0)$.

The first condition implies that when the cell *e* is fully loaded and *k* does not serve any mobile, (i.e. $\chi_e - \chi_k$ approaches 1) it is worth keeping the handover margin $HM(e,k)$ to the lowest value.
The second condition is used to avoid ping pong effect. It implies that when the cell *e* is over loaded and the cell *k* is less-loaded, cell *e* pushes mobiles to cell *k* and conversely cell *k* delays handover to cell *e*.

The function *f* can be approximated by a polynomial (with Taylor series development). The polynomial coefficients can be dynamically determined using learning techniques. The concept of fuzzy reinforcement learning used in an earlier study on 3G-soft-handover optimization could be well suited [6]. In the present study, we restrict the development of *f(x)* to the order 1:

$$f(x) = f(0) + (f(0) - HM_{max})x \qquad (5)$$

The development of order 0 corresponds to the classical case without any auto-tuning. So the simulation aims at comparing the development of order 1, namely with auto-tuning to the case without auto-tuning (denoted as *classical case*).

## 4. Simulations and results

To evaluate the performance of the proposed auto-tuning method, a dynamic simulator developed in *Matlab* tool has been utilized. The simulator performs correlated snapshots to account for the time evolution of the network. At the end of each time step that can typically vary from a tenth to one second, the new mobile positions are updated, new users are admitted and some other users leave the network (end their communications or are dropped) and handover events are processed.
Simulations have been carried out on a LTE network composed of 45 eNBs (figure. 1). Each eNB has a fixed capacity equals to 25 PRBs (corresponding to a 5 MHz bandwidth). A frequency reuse factor equal to 3 is used. The studied scenario uses a non-uniform traffic distribution resulting in some unbalanced cell loads. Only the FTP service class is considered. An FTP call is generated by a Poisson process and the communication duration of each user depends on its bit rate. Each user is allocated at least one PRB and at most 4 PRBs to download a file of 5 Mbytes.
The value of the function *f* in 0 equals 6 dB. The minimum and the maximum handover margin values, $HM_{min}$ and $HM_{max}$, are set respectively to 0 dB and 12 dB.

Figure 2 presents the access probability (the complementary of the blocking rate) versus the traffic intensity for the case of auto-tuning compared with the classic case without auto-tuning. As expected, the gain of using auto-tuning is important when the traffic intensity is low because the dispersion of cell loads is still high. For high traffic intensities, all cell loads approach 1 and the load difference of adjacent cells becomes too small to benefit from traffic balancing. According to the auto-tuning of order 1 and more, the handover margin tends to the default handover margin (*f(0)* in eq (4)) when the traffic increases and all loads tend to 1.

Looking closely to figure 2, 80% of higher traffic intensity can be obtained by the auto-tuning with the same access success rate equals to 95% (7.3 mobiles/s can be accommodated in the auto-tuning case versus only 4 mobiles/s for the other case).

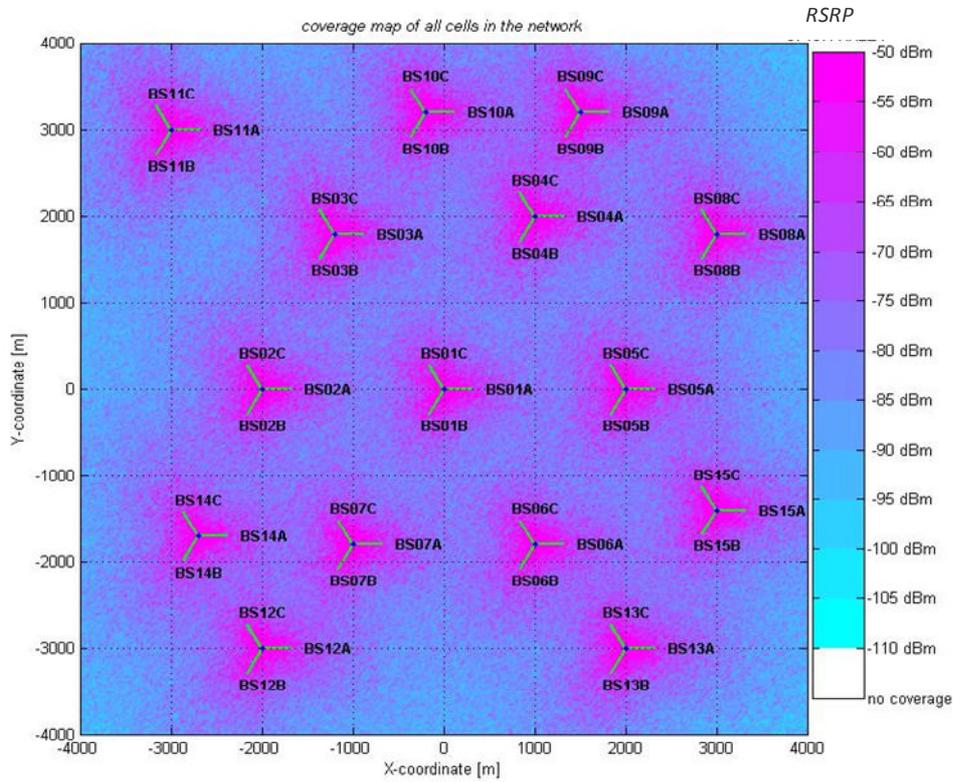

*Fig. 1. The network layout including coverage of each eNB*

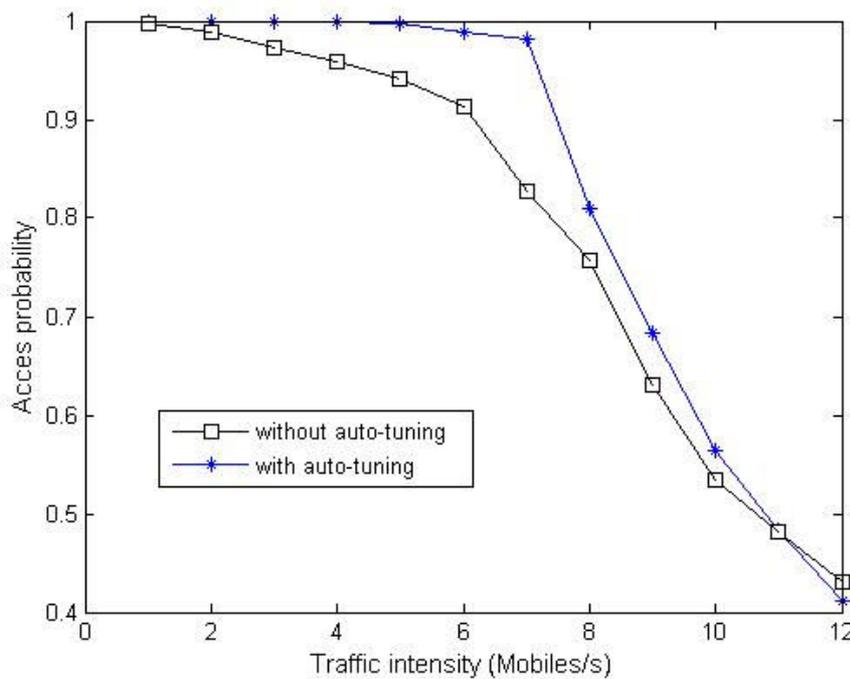

*Fig. 2. Admission probability as a function of the traffic intensity for auto-tuned handover compared with fixed handover margin (6dB).*

In figure 3, we present the connection holding rate (the complementary of the dropping rate) as a function of the traffic intensity. We notice that the variation of the holding rate is small when the traffic increases. This is due to the fact that a mobile is not dropped when there is not enough resources in the serving cell but instead its throughput decreases (i.e. the number of allocated PRBs decreases). The auto-tuning gain for this metric is modest. The trend of the two curves for the high traffic intensity, confirms again that the auto-tuning tends to the classic case in very high traffic condition.

Figure 4 shows the average throughput per user as a function of the traffic intensity. The throughput per user is a decreasing function of the traffic rate since it is an increasing function of the SINR. Here the gain of the auto-tuning is too high. For instance, for the traffic intensity equals to 5 users/s, the throughput per user is approximately 1.15Mbyte/s whereas for the classic case, it is only 0.975Mbyte/s. This gain is explained by two reasons:
1. The implementation of resource allocation: when there are enough resources in the cell, the user gets the maximum number of PRBs. So its bit rate is high and the user ends quickly its communication. As a consequence, it rapidly liberates resources for new users. This explains the gain in successful access rate brought by the auto-tuning.
2. Interference diversity: Due to the auto-tuning, the distribution of inter-cell interference becomes more or less the same in each eNB since the interferences experienced by each user depend on the load of the neighbouring cells. Hence, load balancing leads to interference diversity.

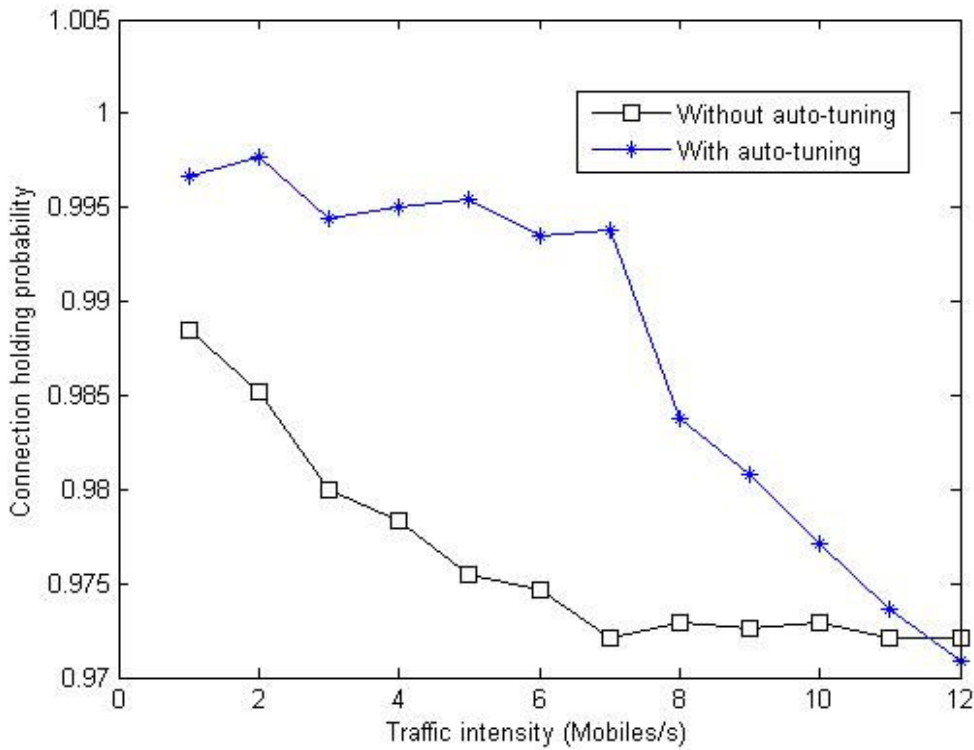

*Fig. 3. Connection holding probability as a function of the traffic intensity for auto-tuned handover compared with fixed handover margin (6dB).*

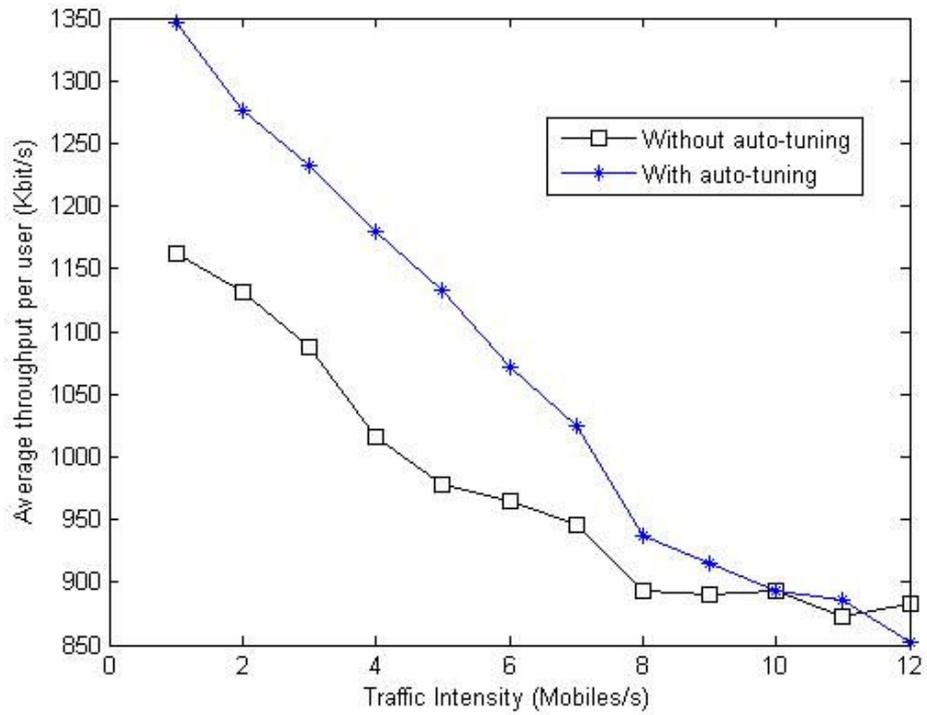

*Fig. 4. Average throughput per user versus the traffic intensity for auto-tuned handover compared with fixed handover margin (6dB).*

In figure 5, the cumulative distribution for SINR is presented for both the auto-tuning and the classical cases. The traffic intensity has been set to 8 mobiles/s. The interference diversity generated by the auto-tuning mechanism leads to an increase of the perceived SINR.

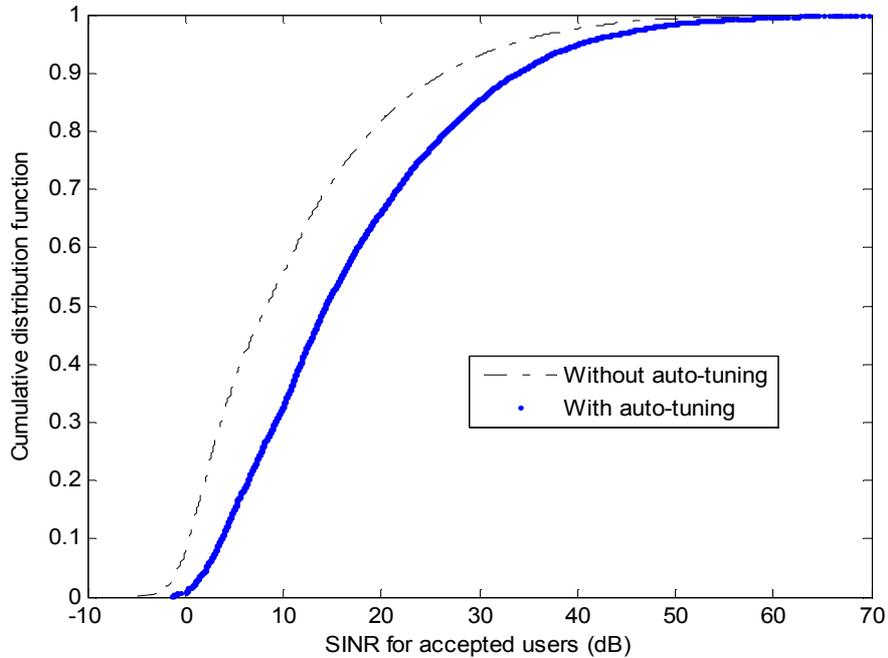

*Fig. 5. Cumulative distribution function of the SINR for network with and without auto-tuning, for traffic intensity equals to 8 mobiles/s.*

## 5. Conclusion

This paper has investigated the auto-tuning of mobility algorithm in LTE system. The mobility is based on hard handover. The hard handover margin involving each couple of eNB governs the hard handover and its value directly affects the radio load distribution between cells. The auto-tuning of the handover margin parameters balances the traffic between neighbouring cells. As a consequence, the system capacity is increased and the user perceived quality of service, namely the user throughput, is enhanced. The auto-tuning process has been implemented in a dynamic system level simulation of a non regular (e.g. cells are not hexagonal) LTE network. Significant improvement in SINR has been obtained, and more than 15% increase in user throughput has been achieved by the mobility auto-tuning.